\documentclass[aps,superscriptaddress,reprint]{revtex4-1}
\PassOptionsToPackage{warn}{textcomp}
\usepackage{times}
\usepackage[ngerman,english]{babel}
\usepackage{upgreek}
\usepackage{units}
\usepackage{graphicx}
\usepackage{amsmath}
\usepackage{amssymb}
\usepackage{amsthm}
\usepackage{mathcomp}
\usepackage{pgffor}
\usepackage{mathrsfs}
\usepackage{braket}
\usepackage{dsfont}
\usepackage{esvect}
\usepackage{esdiff}
\usepackage{float}
\usepackage{url}
\usepackage{framed}
\usepackage{xcolor, xspace}
\usepackage{ulem}
\usepackage[displaymath,mathlines]{lineno}

  \usepackage{booktabs}

\newcommand*\patchAmsMathEnvironmentForLineno[1]{%
  \expandafter\let\csname old#1\expandafter\endcsname\csname #1\endcsname
  \expandafter\let\csname oldend#1\expandafter\endcsname\csname end#1\endcsname
  \renewenvironment{#1}%
     {\linenomath\csname old#1\endcsname}%
     {\csname oldend#1\endcsname\endlinenomath}}%
\newcommand*\patchBothAmsMathEnvironmentsForLineno[1]{%
  \patchAmsMathEnvironmentForLineno{#1}%
  \patchAmsMathEnvironmentForLineno{#1*}}%
\AtBeginDocument{%
\patchBothAmsMathEnvironmentsForLineno{equation}%
\patchBothAmsMathEnvironmentsForLineno{align}%
\patchBothAmsMathEnvironmentsForLineno{flalign}%
\patchBothAmsMathEnvironmentsForLineno{alignat}%
\patchBothAmsMathEnvironmentsForLineno{gather}%
\patchBothAmsMathEnvironmentsForLineno{multline}%
}

\usepackage{multibib}
\newcites{sec}{References Methods}

\newcommand*\downk{\ket{\downarrow}}

\newcommand*\upk{\ket{\uparrow}}

\newcommand*\ee{\mathrm{e}} 
\newcommand*\ii{\mathrm{i}}

\begin{document}
\title{Motional Fock states for quantum-enhanced amplitude and phase measurements with trapped ions}
\author{Fabian Wolf} \author{Chunyan Shi} \author{Jan C. Heip}

\affiliation{ Physikalisch-Technische Bundesanstalt, 38116 Braunschweig,
Germany }
\author{Manuel Gessner}\author{ Luca Pezz\`e} \author{Augusto Smerzi}
\affiliation{QSTAR, INO-CNR and LENS, Largo Enrico Fermi 2, I-50125 Firenze, Italy}
\author{Marius Schulte} \author{Klemens Hammerer}
\affiliation{Institute for Theoretical Physics, Institute for Gravitational Physics (Albert Einstein Institute), Leibniz Universit\"at Hannover, Appelstrasse 2, 30167 Hannover, Germany} 
\author{Piet O. Schmidt} 
\affiliation{Physikalisch-Technische Bundesanstalt, 38116 Braunschweig, Germany }
\affiliation{Institut f\"ur Quantenoptik, Leibniz Universit\"at Hannover, 30167 Hannover, Germany}

\noindent
\begin{abstract}
Non-vanishing fluctuations of the vacuum state are a salient feature of quantum theory. 
These fluctuations fundamentally limit the precision of quantum sensors.
Nowadays, several systems such as optical clocks~\cite{ludlow_optical_2015}, gravitational wave detectors~\cite{ligo_scientific_collaboration_and_virgo_collaboration_observation_2016}, matter-wave interferometers, magnetometers~\cite{wasilewski_quantum_2010}, and optomechanical systems~\cite{aspelmeyer_cavity_2014} approach measurement sensitivities where the effect of quantum fluctuations sets a fundamental limit, the so called standard quantum limit (SQL). 
It has been proposed~\cite{caves_quantum-mechanical_1981} that the SQL can be overcome by squeezing the vacuum fluctuations. 
Realizations of this scheme have been demonstrated in various systems~\cite{gross_nonlinear_2010,vahlbruch_detection_2016, aasi_enhanced_2013,leroux_orientation-dependent_2010,hosten_measurement_2016,ockeloen_quantum_2013,kienzler_quantum_2015}. 
However, protocols based on squeezed vacuum crucially rely on precise control of the relative orientation of the squeezing with respect to the operation imprinting the measured quantity. Lack of control can lead to an amplification of noise and reduces the sensitivity of the device.
Here, we experimentally demonstrate a novel quantum metrological paradigm based on phase insensitive Fock states~\cite{pezze_ultrasensitive_2013} of the motional state of a trapped ion, with applications in frequency metrology and displacement detection.
The measurement apparatus is used in two different experimental settings probing non-commuting observables with sensitivities beyond the SQL.  In both measurements, classical preparation and detection noise are sufficiently small to preserve the quantum gain in a full metrological protocol. 
\end{abstract}

\maketitle

Advances in the ability to control quantum systems together with the suppression of classical noise originating from technical imperfections, has led to the emergence of sensors that are limited in their performance by quantum noise. For more than thirty years it has been known that certain types of non-classical states can reduce the effect of quantum noise and thus enhance the sensitivity of measurement devices beyond the classical limit~\cite{caves_quantum-mechanical_1981}. Taking advantage of this sub-SQL sensitivity requires not only the preparation of the non-classical state with high fidelity, but also the prevention of signal loss in the entire measurement protocol. This has been achieved e.g. with squeezed states and Schr\"odinger-cat or N00N states in interferometric settings ~\cite{aasi_enhanced_2013,leroux_orientation-dependent_2010,hosten_measurement_2016,ockeloen_quantum_2013}. A common restriction of these types of non-classical states is the need for control over the relative phase between the state creation and the measurement interaction~\cite{hempel_entanglement-enhanced_2013,lo_spin-motion_2015}. In a phase-space picture, squeezing along the displacement direction enhances the sensitivity for amplitude measurements, but weakens the sensitivity for phase measurements. 
Here we present sub-SQL measurements of amplitude and phase of the motional state of a trapped ion using the same motional Fock state. For this purpose, we operate the sensing device in two different experimental settings. Firstly, the amplitude of the ion’s oscillation is varied and the phase is kept constant, which realizes a displacement or force sensor~~\cite{shaniv_quantum_2017,biercuk_ultrasensitive_2010,gilmore_amplitude_2017}. Secondly, the Fock state is displaced with a fixed amplitude in a Ramsey-like interferometry sequence to measure the phase of the ion’s oscillation which implements a measurement of the oscillation frequency of the ion in the trap.
Furthermore, we prove that the non-classicality in terms of the Glauber-Sudarshan P-function is the resource for the metrological gain ~\cite{rivas_precision_2010} and that Fock states are optimal for sensing displacements with unknown phase. 
\begin{figure*}
 \centering
 \includegraphics[width=183mm]{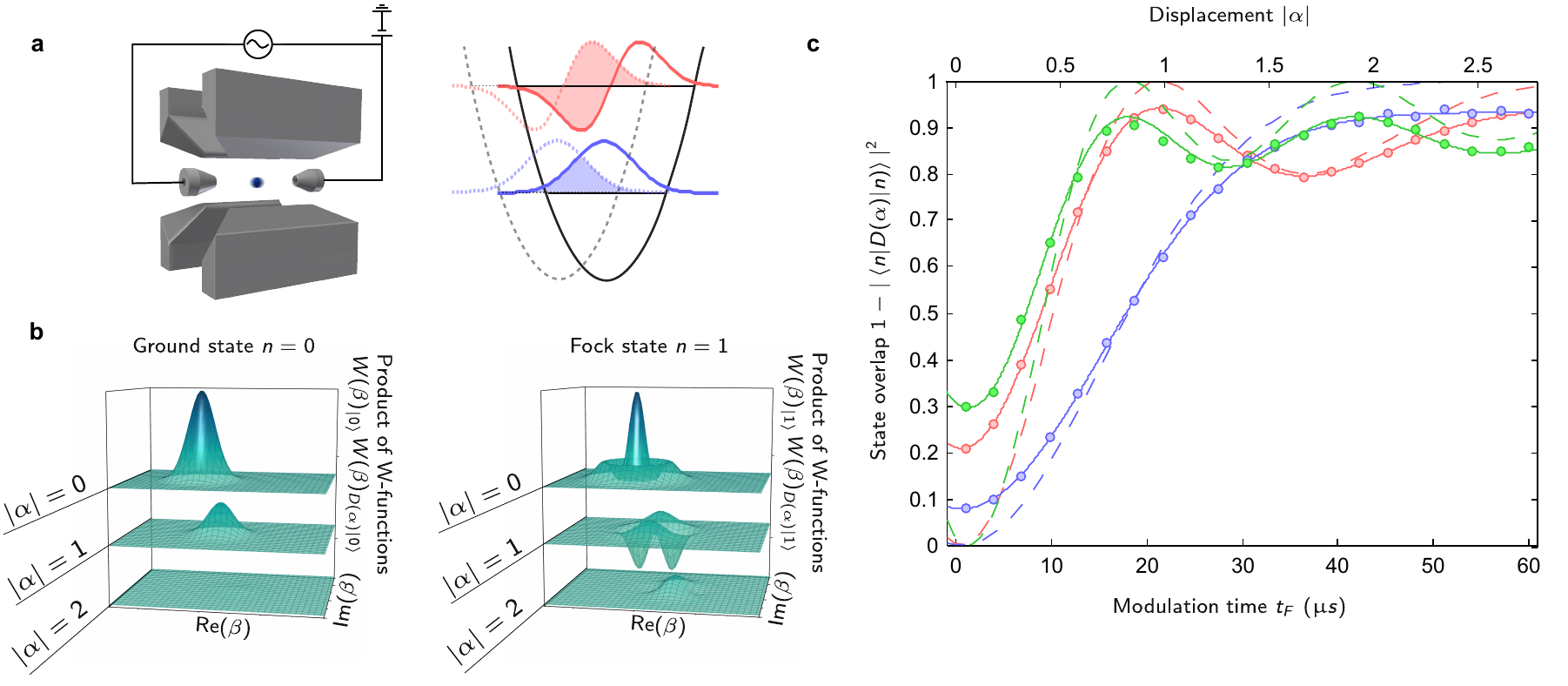}
 \caption{\textbf{Measuring displacement amplitudes.}
  \textbf{\textsf{a}}
  The left panel shows a schematic of the experimental setup: 
  A magnesium ion is trapped in a linear Paul trap and an additional ac voltage on the end electrodes implements the motional displacement(coherent  excitation of motion). 
  On the right panel the initial motional state wave functions are shown as dashed lines ($n=0$ blue, $n=1$ red). 
  After applying a resonant oscillating force, the ion's motion is in a displaced Fock state whose wave function is depicted by the solid colored lines. 
  To infer the displacement, the wave function overlap is measured as sketched by the colored  areas.
  \textbf{\textsf{b}} 
  Illustration of the mechanism behind enhanced sensitivity of Fock states to displacements. The left and the right panel show the theoretically calculated product of the initial and final Wigner function  for three different displacements for the Fock states $n=1$ and $n=0$, respectively.
  By integrating over the whole phase space it is possible to infer the state overlap shown in (\textsf{c}). 
  The negative parts of the $n=1$ Fock states lead to a vanishing integral for $\alpha=1$, before the contours of the Wigner functions are fully separated. 
  For states without negative regions in the Wigner function, such as Gaussian states, the integral can only vanish when the product of Wigner functions vanishes over the whole phase space.
  \textsf{\textbf{c}} 
  The graph shows the outcome of the state overlap measurement for three different initial Fock states (blue: $n=0$, red: $n=1$, green: $n=2$). 
  The oscillating force is applied for different duration $t_\mathrm{F}$.
  The solid curves are fits of the equation given in the main text to the data and the dashed lines are the corresponding theoretical curves assuming full contrast.
  The fit for the motional ground state is used to calibrate the upper $x$-axis, denoting the displacement amplitude $|\alpha|$. 
  Error bars for the standard error of the mean (s.e.m.) due to quantum projection noise are too small to be seen. 
  Each point is an average of approximately $10\,000$~experiments.}
   \label{Fig:Displacement}
\end{figure*}

The experiments are performed with a single $^{25}$Mg$^+$ ion
confined in a linear Paul trap.  Excited motional Fock states are created
starting from the motional and electronic ground state~\cite{wan_efficient_2015}, through a sequence of laser-driven blue and red
sideband pulses that each add a quantum of motion while changing the internal
state of the ion~\cite{meekhof_generation_1996,wineland_experimental_1998}.  A
calibrated displacement $\hat{D}\left( \alpha \right)=\exp{(\alpha \hat{a}^\dagger-\alpha^{*}\hat{a})}$ is implemented by exposing the
ion to an electric field oscillating at the trapping frequency of
$\omega_z=\unit[2\pi\times1.89]{MHz}$.
The displacement amplitude $|\alpha|$ can be controlled through the modulation time
$t_{\mathrm{F}}$ (see Methods for more
details). It is measured by mapping the overlap between initial and displaced state
onto the atomic qubit ($\left|\uparrow \right\rangle$, $\left|\downarrow\right\rangle$, encoded in two hyperfine states of the
$^2$S$_{1/2}$ electronic ground state of $^{25}$Mg$^+$), where state-readout is
performed using state dependent fluorescence~\cite{hemmerling_novel_2012}.  The mapping
process is implemented by a sequence  of sideband rapid adiabatic passage (RAP)~\cite{gebert_detection_2016} and microwave pulses and is described in more
detail in the Methods section.

\begin{figure*} \centering
  \includegraphics[width=183mm]{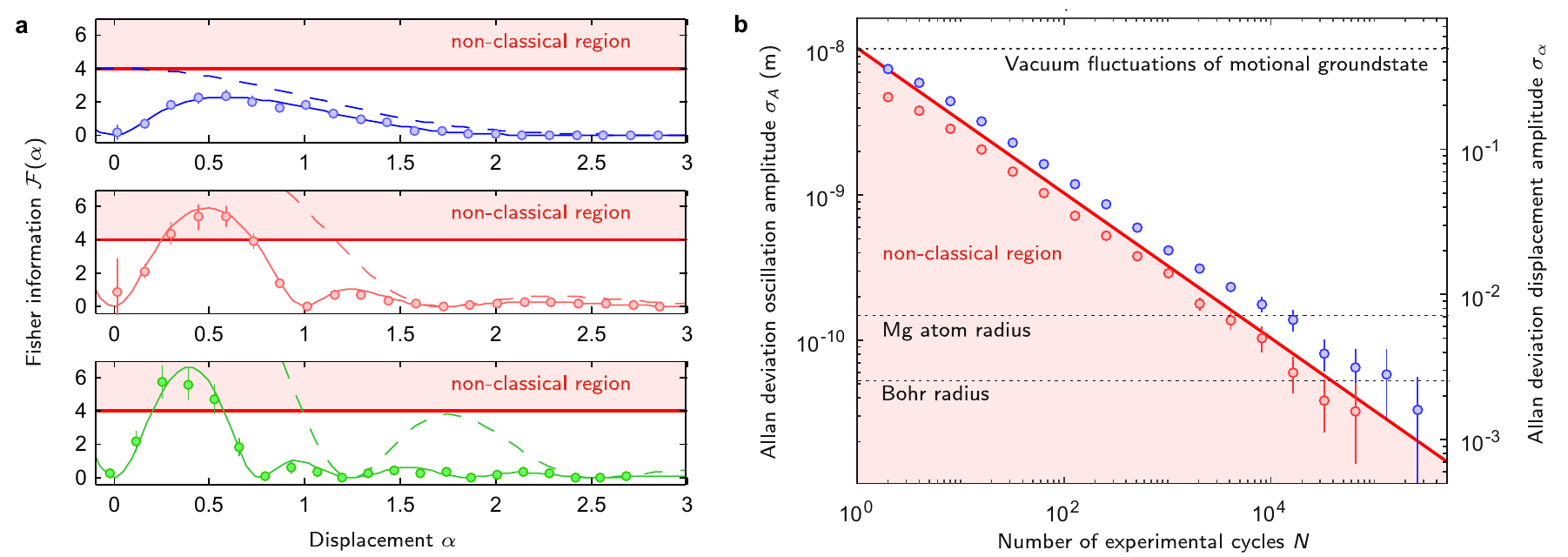}
  \caption{
    \textbf{Evaluation of displacement measurement.}
  \textbf{\textsf{a}} 
  Fisher information extracted from the amplitude
  measurement for three different initial Fock states (blue: $n=0$, red: $n=1$,
  green: $n=2$). The dashed lines show the theoretically expected value for full
  contrast, whereas the solid lines show the expected values considering the
  reduced contrast and offset (extracted from fit, see Fig,~\ref{Fig:Displacement}). The standard
  quantum limit is depicted by the red line. The Fock states with $n>0$ significantly
  surpass this limit. The error bars indicate the standard deviation (see
  supplementary information for more details).
  \textbf{\textsf{b}}
  Allan deviation for an amplitude  measurement around $\alpha=0.59$  with a coherent state (blue circles) and
  with a $n=1$ displaced Fock state (red circles). The solid line shows the
  quantum projection noise limit from Eq.~\ref{eq:SQL} with the classical theoretical optimum $\mathcal{F}_{n=0}=4$. } 
\label{Fig:FisherInfo}
\end{figure*}

Fig.~\ref{Fig:Displacement} \textbf{\textsf{a}} shows the principle and
Fig.~\ref{Fig:Displacement} \textbf{\textsf{c}} the result of such a measurement for three different initial Fock states
($n=0,1,2$).  The expected state overlap is given by
$\left|\left\langle n|D(\alpha)|n \right\rangle\right|^2=\exp(-|\alpha|^2)\left(\mathcal{L}_n(|\alpha|^2)\right)^2$,
with the Laguerre polynomials $\mathcal{L}_n$~\cite{de_oliveira_properties_1990}.
The measurement suffers from reduced contrast due to imperfections in state
preparation and the detection process, which are of technical nature and pose no fundamental limitation. To account for these imperfections the fitting function depicted by the solid line in Fig.~1\textbf{\textsf{c}} is 
$P_\mathrm{fit}=C_1+C_2\exp(-|\dot{\alpha}t_{\mathrm{F}}|^2)\left(\mathcal{L}_n(|\dot{\alpha}t_{\mathrm{F}}|^2)\right)^2,$
  \label{eq:fitfunction}
with free parameters $C_1$,  $C_2$, and $\dot{\alpha}$. The fitted value of
$\dot{\alpha}$ for the $n=0$ data is used to calibrate the displacement
strength shown on the upper $x$-axis. 

In contrast to the monotonous behavior of the $n=0$ measurement outcome,
the data for the excited Fock states exhibit fringes due to interference in phase space ~\cite{schleich_oscillations_1987}. The interference fringes and the resulting metrological gain can be understood as a consequence of the negative regions of the Wigner function as shown in Fig.~\ref{Fig:Displacement} \textbf{\textsf{b}}. 
In phase space the overlap of two quantum states is represented by the integral
over the product of the Wigner functions 
\begin{eqnarray}
\left| \left\langle \psi_\mathrm{i}|\psi_\mathrm{f}\right\rangle\right|^2=\int\!\!\!\!\int\!\! \mathrm{d \beta}\,W(\beta)_{\left|\psi_\mathrm{i}\right\rangle}W(\beta)_{\left|\psi_\mathrm{f}\right\rangle}. 
\end{eqnarray}
In consequence the overlap between a classical state (with positive Wigner
function) and its displaced counterpart  only vanishes for vanishing overlap of
the phase space contours of the involved states (see left panel in
Fig.~\ref{Fig:Displacement} \textbf{\textsf{b}}). However, if
the quantum state reveals negative values in the Wigner function, as is the
case for Fock states, the negative parts in the product can cancel the positive
parts and lead to vanishing overlap before the wave packets are spatially
separated (see right panel in Fig.~\ref{Fig:Displacement} \textbf{\textsf{b}}). 
The metrological gain can be quantified
by the Fisher information $\mathcal{F}$ for the displacement
measurement which can be extracted from the data shown in Fig.~\ref{Fig:Displacement}~(see Methods for details). The result is shown in Fig.~\ref{Fig:FisherInfo} \textsf{\textbf{a}}. 
For a displacement of $\alpha=0.59$ the measured Fisher information for the n=1 Fock state measurement is
$\mathcal{F}_{n=1}=5.37(63)$ (error is standard deviation (s.d)) which implies a metrological gain of
$g_{SQL}={\frac{\mathcal{F}_{n=1}(\alpha=0.59)}{\mathcal{F}_{SQL}}}=\unit[1.3]{dB}$
compared to the theoretical SQL and
$g={\frac{\mathcal{F}_{n=1}(\alpha=0.59)}{\max_\alpha(\mathcal{F}_{n=0})}}=\unit[3.6]{dB}$
compared to the achieved performance for the $n=0$ state
($\mathcal{F}_{n=0}(\alpha=0.59)=2.36(30)$).  
This corresponds to a reduction in averaging time by more than a factor of two for the same displacement resolution. 
The Fisher information is directly
linked to the achievable measurement uncertainty by the  Cram\'er-Rao  bound
\begin{equation} 
  \Delta\alpha \geq \Delta{\alpha}^{\mathrm{CR}}=\frac{1}{\sqrt{N \mathcal{F} (\alpha)}}, 
  \label{eq:SQL}
\end{equation}
where $N$ is the number of
independent experimental cycles. In agreement with the  Cram\'er-Rao  bound,
the uncertainty for the displacement measurement shown in
Fig.~\ref{Fig:FisherInfo}~\textbf{\textsf{b}} in the form of an Allan deviation $\sigma_\alpha$ averages down faster for the
n=1 Fock state (red circles) compared to the ground state (blue circles). Note that for white noise, the Allan deviation $\sigma_\alpha$ and standard deviation $\Delta \alpha$ are identical.
The achieved sensitivity for displacement of $\sigma_A(N=2^{16})=\unit[65(23)]{pm}$ for $n=0$ and $\sigma_A(N=2^{16})=\unit[32(18)]{pm}$ for $n=1$ can be translated into force measurement resolution~(see Methods) of $\unit[1.8(0.6)]{yN}$ for $n=0$ and $\unit[0.9(0.5)]{yN}$ for $n=1$ after $N=2^{16}=65\,536$ independent experiments, where an experimental cycle takes $\unit[8.1]{ms}$ and $\unit[9.5]{ms}$ for the $n=0$ and $n=1$ measurement, respectively. 

For pure states it can be shown that the quantum Fisher information for a
displacement along phase space quadrature $\hat{R}(\phi_\mathrm{LO})=\left(\sin\left(\phi_\mathrm{LO}\right)\hat{X}+\cos\left(\phi_\mathrm{LO}\right)\hat{P}\right)/\sqrt{2}$ is proportional to the variance of the conjugate variable~\cite{braunstein_statistical_1994,pezze_non-classical_2016}, i.e. $\mathcal{F}_Q=16\left(\Delta\hat{R}{(\phi_\mathrm{LO}+\pi/2)}\right)^2$, where $\phi_\mathrm{LO}$ is the phase of the oscillating force. 
According to this relation the metrological gain can be understood as a consequence of the antisqueezing in the generating, conjugate quadrature. The fact that Fock states are antisqueezed along all directions provides an illustrative explanation for the their phase insensitive gain. 
For a Fock state of order $n$, the quantum Fisher information is given by $\mathcal{F}_Q=8n+4$.
In general, any enhancement beyond the classical limit can be traced back to non-classicality as defined in terms of the Glauber-Sudarshan P-distribution. Furthermore, for displacement measurements with an unknown phase, it can be shown that Gaussian states, or mixtures thereof, will always perform suboptimally, whereas Fock states are indeed optimal~(see supplementary information).

As a consequence of the insensitivity of the Fock state to the displacement direction, the same state can be
employed for quantum-enhanced sensing of displacement amplitude \textit{and} phase changes. 
We demonstrate this feature by measuring the oscillation frequency of the trapped ion with sub-SQL resolution in a Ramsey-like experiment as sketched in the inset of Fig.~\ref{Fig:Ramsey} (see supplemental information and Extended Data 
Fig.~\ref{Fig:RamseyFull}
for more detailed information).
The Ramsey sequence starts with the
initialization of the ion's motion in a Fock state~(\textsf{\textbf{I}}) and a subsequent
displacement in phase space~(\textsf{\textbf{II}}). 
If the drive for the displacement was detuned by $\delta$ from the trap
frequency, the displaced state will evolve in phase space on a circle around the origin and accumulate a phase $\phi=\delta\times T$ compared to the driving
field during the waiting time $T$~(\textsf{\textbf{III}}). Undoing the displacement~(\textsf{\textbf{IV}}) maps this phase onto a
residual displacement $\delta \alpha$ that can be detected with the overlap
detection method introduced above. The center fringe of the Ramsey pattern for waiting time
$T=\unit[50]{\upmu s}$ and initial displacement $\alpha=1.6$ is shown in Fig.~\ref{Fig:Ramsey}. As 
illustrated by the data shown in the inset, the width of the center fringe
decreases with increasing Fock state order. The full-width-half-maximum~(FWHM)
is extracted from a Gaussian fit to the center peaks. Note that a narrower width does not necessarily imply a metrological gain. For an increase in Fisher information the slope of the line feature has to increase. For n=2 the reduction in linewidth is fully compensated by the reduced contrast. The whole Ramsey pattern
for the different initial Fock states is shown in Extended Data Fig.~\ref{Fig:RamseyFull} and
the theoretical line shape is derived in the supplementary information. 

\begin{figure}
  \centering 
  \includegraphics[width=89mm]{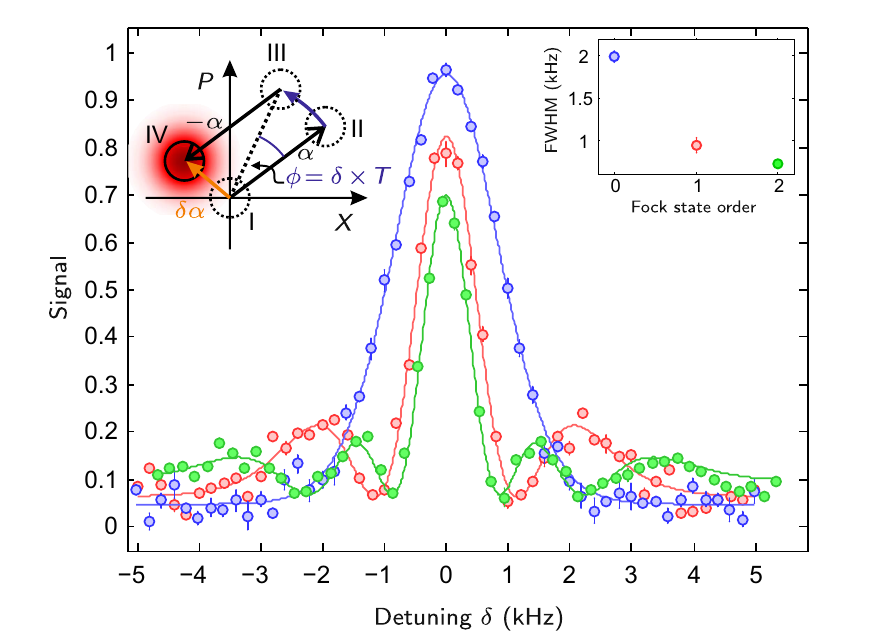}
  \caption{
    \textbf{Trapping frequency measurement.}
    Center fringe of the Ramsey pattern for three different initial Fock
    states (blue: $n=0$, red: $n=1$, green: $n=2$). The circles show the
    experimental data, which is the population probability of the ion in the
    $\left|\downarrow\right\rangle$-state ($\left| \uparrow\right\rangle$ for $n=0$, see text for details). The solid lines
    show a fit of the theoretically expected lineshape to the data (see
    supplementary information). The inset shows the full-width-half-maximum
    (FWHM) extracted from a Gaussian fit to the data. Each data point consists
    of 1000 experimental cycles, evaluated with a distribution fit
    technique~\cite{hemmerling_novel_2012} } 
    \label{Fig:Ramsey} 
 \end{figure}

\begin{figure} \centering
  \includegraphics[width=89mm]{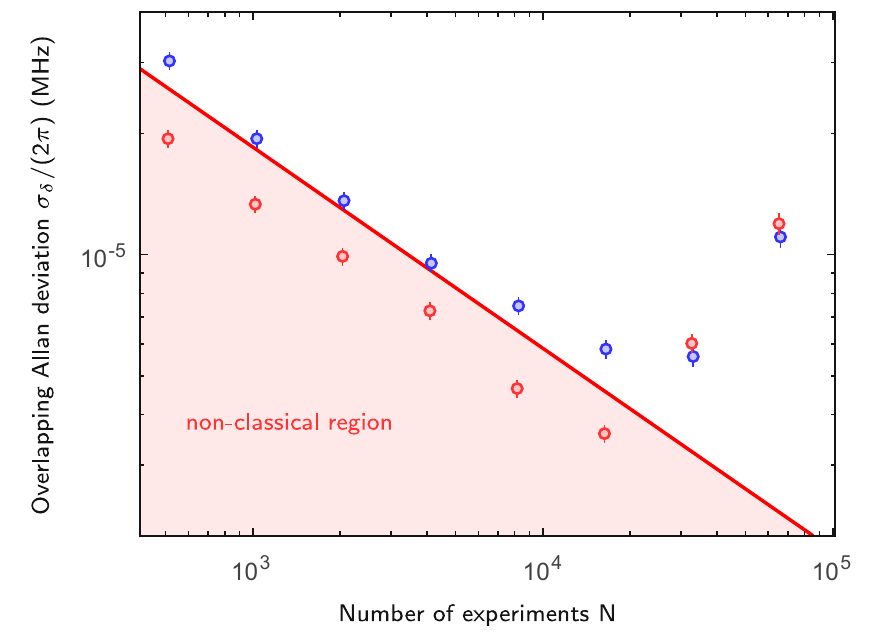} \caption{\textbf{Evaluation of trapping frequency measurement.} Overlapping Allan
  deviation for frequency measurement with a coherent state (blue circles) and
with a $n=1$ displaced Fock state (red circles). The solid line shows the
quantum projection noise limit from Eq.~\ref{eq:SQLFreqAllan}.}
\label{Fig:FreqAllan}
\end{figure}

To evaluate the performance of the quantum sensing techniques, we have performed
a trapping frequency measurement by two-point sampling and analyzed the data in terms of an Allan deviation (see
Fig.~\ref{Fig:FreqAllan}). Since the $n=2$ Fock
state in our case does not provide an additional metrological advantage (see
Fig.~\ref{Fig:FisherInfo}) as a consequence of the reduced contrast caused by technical
limitations of the implementation, we have performed the Allan deviation analysis for the
$n=0$ and $n=1$ Fock state only.
The measurement has been performed in an interleaved pattern with an average cycle time of $\unit[6.6]{ms}$ and $\unit[7.8]{ms}$ for the $n=0$ and $n=1$ measurement, respectively.
The Allan deviation for the $n=0$ protocol averages down to
$\sigma_\delta^{n=0}=\unit[2\pi\times5.8(3)]{Hz}$. The achievable resolution is limited
by a linear drift of the trapping frequency, which leads to an increase in the
Allan deviation for long averaging times. The red line in
Fig.~\ref{Fig:FreqAllan} is the SQL given by 
\begin{equation}
  \sigma_\delta^{\mathrm{SQL}}=\frac{1}{2\left(T+t_F\right)}\frac{1}{|\alpha|\sqrt{N}}\,,
  \label{eq:SQLFreqAllan}
\end{equation}
which is the lowest statistical uncertainty achievable with a classical state.
For the quantum-enhanced technique with $n=1$, the overlapping Allan deviation reaches
$\sigma_\delta^{n=1}=\unit[2\pi\times 3.6(2)]{Hz}$ before it increases due to the linear drift.
Using the $n=1$ Fock state improves the frequency resolution by more than $\unit[60]{\%}$ compared to the vacuum state. This is a direct consequence of the quantum-enhanced reduction in averaging time, which allows measuring the trapping frequency with high accuracy before it starts drifting.

In summary, we have demonstrated a quantum-enhanced sensing scheme based on
motional Fock states to measure the amplitude and the phase of an oscillating
force with sensitivities below the standard quantum limit. This is complementary to another quantum enhanced method to sense motional frequencies based on phase-sensitive superpositions of Fock states~\cite{mccormick_private_2018}. 
The sensing scheme does not require any phase relation between the displacement and the quantum state of the detector, which is an important feature when measuring arbitrary interactions without prior phase information.
In contrast to the phase-insensitive schemes exploiting correlated modes of atomic ensembles~\cite{lucke_twin_2011,wasilewski_quantum_2010}, our scheme requires no mode entanglement.
  
Quantum logic spectroscopy~\cite{schmidt_spectroscopy_2005} based on motional
displacements~\cite{wan_precision_2014,hume_trapped-ion_2011} will benefit from the
presented amplitude detection technique, in particular for state detection and
spectroscopy of non-closed transitions~\cite{wolf_non-destructive_2016}, where
scattering on the spectroscopy ion has to be reduced to a minimum. 

The presented quantum-enhanced frequency measurement can help to further improve
high precision mass measurements of atoms in Paul traps~\cite{staanum_sympathetically-cooled_2010} and g-factor measurements of subatomic particles, such as (anti-)protons in Penning traps~\cite{sturm_high-precision_2014,ulmer_high-precision_2015}. 
Both cases will benefit from a quantum logic approach, in which a mass or spin-dependent force on the particle of interest is probed with quantum-enhanced sensitivity by a nearby well-controllable logic ion using motional Fock states.
 Further improvements in sensitivity can be achieved by employing techniques that allow the generation and overlap detection of larger Fock states with high fidelity. Scalable overlap measurements for Fock states up to
$n=10$ have been reported~\cite{ding_cross-kerr_2017}, allowing phase-insensitive
suppression of quantum projection noise of up to $\unit[13.2]{dB}$.

\paragraph{Acknowledgements}
We acknowledge support from the DFG through CRC~1227 (DQ-mat), projects A06 and B05, and
the state of Lower Saxony, Hannover, Germany. M.G. acknowledges support by the Alexander von Humboldt foundation.

\paragraph{Author Informations} The authors declare that they have no
 competing financial interests. Correspondence and requests for materials
should be addressed to P.O.S. \\(email: piet.schmidt@quantummetrology.de).
\onecolumngrid
\clearpage
\section*{methods}
\renewcommand{\figurename}{Extended Data Figure}
\setcounter{figure}{0}

\renewcommand{\figurename}{Extended Data Figure}
\section{Trap modulation to implement displacement operator}
Applying a resonantly oscillating electric field at the position of the ion leads to a
displacement of the ion's motional state in phase
space~\cite{carruthers_coherent_1965}. The interaction Hamiltonian for a trapped
ion with an additional time dependent potential $V(t,z)=-qE(t)\hat{z}$, where $q$ and
$\hat{z}$ are the charge and the position of the ion, respectively and $E(t)$ is the time dependent
electric field, that is assumed to be spatially constant over the extent of the ion's wave function, can be written as
\begin{equation}
 \hat{H}=-qE(t)z_0\left(\hat{a}\ee^{-\ii\omega_zt}+\hat{a}^\dagger\ee^{\ii\omega_zt}\right)\,,
\end{equation} 
in an interaction picture with respect to the free harmonic oscillation Hamiltonian $\hat{H}_\text{
HO}=\hbar\omega_z \hat{a}^{\dagger}\hat{a}$ and $\hat{z}$ is the position operator 
$\hat{z}=z_0\left(\hat{a}\ee^{-\ii\omega_zt}+\hat{a}^\dagger\ee^{\ii\omega_zt}\right)$ with
the annihilation(creation) operator $\hat{a}(\hat{a}^{\dagger})$ and ground state wave
function extent $z_0=\sqrt{\hbar/2m\omega_z}$.  For an electric field oscillating
at the trapping frequency $\omega_z$, this leads to the static Hamiltonian 
\begin{equation}
 \hat{H}=-\frac{qE_0z_0}{2}\left(\hat{a}\ee^{-\ii \phi_\text{LO}}+\hat{a}^\dagger\ee^{\ii\phi_\text{LO}}\right)\,, 
\end{equation}
where fast oscillating terms (at twice the trapping frequency) are neglected within the rotating wave approximation.
Here, $\phi_\text{LO}$ and $E_0$ are the phase and amplitude of the driving
field, respectively.  The unitary evolution according to this Hamiltonian is
\begin{eqnarray}
 \hat{U}(t)=\ee^{-\frac{\ii}{\hbar}\hat{H} t}=\hat{D}(\alpha)
\end{eqnarray}
 and can be identified as the displacement $\hat{D}(\alpha)=\ee^{\alpha \hat{a}^\dagger-\alpha^* \hat{a}}$ operator
with displacement amplitude $\alpha=\frac{\ii q E_0 z_o}{2\hbar} \ee^{\ii
  \phi_\text{LO}}\times t$.

 \section{Overlap measurement}\label{sec:overlap} 
\begin{figure*} \centering
  \includegraphics[width=183mm]{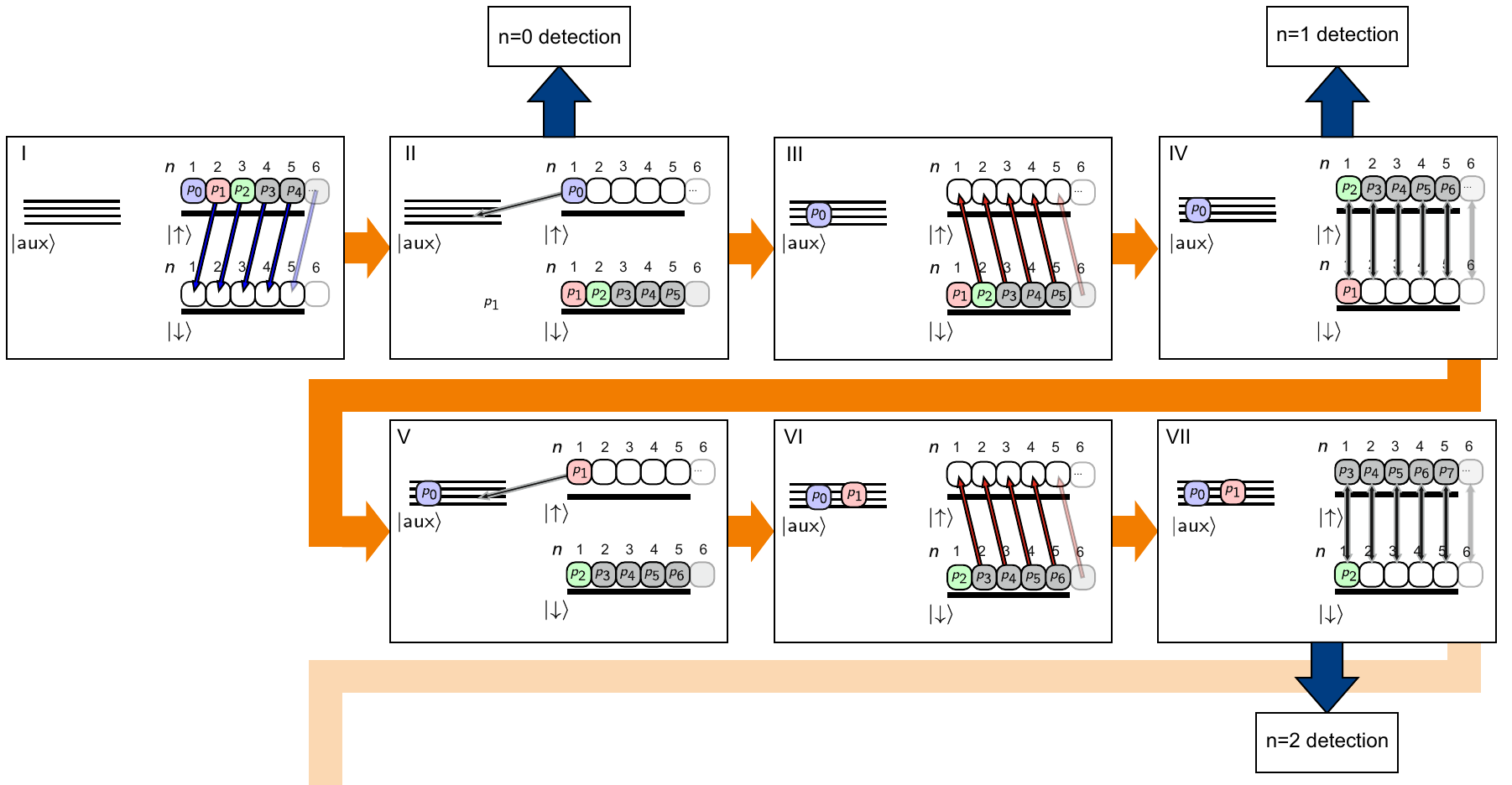}
  \caption{Experimental scheme to measure the motional state overlap with Fock
    state $n$. The reduced level scheme shows the spin states $\downk$, $\upk$
    and the manifold of auxiliary states $\ket{\text{aux}}$ (see text for details). The
    boxes indicate the current motional state within the sequence. The values $p_n$ denote the initial population of the Fock state $n$.
    The sequence starts with the ion in the $\upk$ state and an unknown motional state
    distribution.\textbf{\textsf{(I)}} A blue sideband RAP pulse flips the spin
    and removes a phonon from all excited Fock states, leaving the ground state population untouched.
    $\textsf{\textbf{(II)}}$ The motional ground state population is measured via spin state-selective fluorescence. To detect the population of higher order Fock states, the
    ground state population is hidden by a sequence of rf-pulses in one of the
    auxiliary states $\ket{\text{aux}}_1$. $\textsf{\textbf{(III)}}$ A red sideband RAP
    pulse flips the spin and removes a phonon from all excited Fock states. Therefore only the population
    that was initially in the $n=1$ Fock state remains in the bright spin state
    $\downk$. $\textsf{\textbf{(IV)}}$ Detection of the ions spin gives the
    $n=1$ Fock state population. Alternatively, step
    $\textsf{\textbf{(II)}}$-$\textsf{\textbf{(IV)}}$ can be repeated after a
    rf carrier $\pi$-flop to detect higher order Fock states (see
    $\textsf{\textbf{(V)}}$-$\textsf{\textbf{(VII)}}$).   }
    \label{SupFig:detection} \end{figure*}

All measurements described in the manuscript rely on the ability to measure the
motional state population in a given Fock state. To achieve this, we have
implemented a sequence that transfers a selected initial population $p_n$ to the motional and electronic ground state, while all other motional population is in the $\upk$ state. State-selective fluorescence then provides the population $p_n$. The sequence for measuring $p_0$, $p_1$ and $p_2$  is shown in Fig.~\ref{SupFig:detection}. 
The ion is initialized in the $\upk$-state. 
At the beginning of the detection sequence the motional population $\{p_n\}$ is distributed over several motional Fock states $n$.
(\textsf{\textbf{I}}). A blue sideband rapid adiabatic passage pulse~(RAP) transfers the
internal state to $\downk$, while simultaneously taking out a quantum of motion,
therefore keeping the ground state population untouched~\cite{gebert_detection_2016}.  Averaging the number of $\downk$ and $\upk$ detection events 
after this mapping step provides the $n=0$ population.   For higher order Fock
state detection the protocol has to be extended as follows. The ground state population can
be hidden in a dark auxiliary state $\ket{\text{aux}}$ by radio frequency
pulses~(\textsf{\textbf{II}}). In $´^{25}$Mg$^+$ the Zeeman substates with
$m_F=1,0,-1,-2$ of the $F=2$ dark hyperfine state can be used for this purpose.
A second sideband RAP pulse~(\textsf{\textbf{III}}), this time on the red sideband, flips the spin for all motional states except for the ground state, which
stores the information about the initial Fock state $n=1$ population.
Fluorescence detection of the ion's spin will give the initial $n=1$ Fock state
population. To detect even higher Fock states, the spin is flipped
independent of the motional state to initialize the $\upk$-state again
(\textsf{\textbf{IV}}). Now steps  (\textsf{\textbf{II}}) to
(\textsf{\textbf{IV}}) are repeated until the desired Fock state population
is isolated in the $\downk$ state from the rest of motional population (e.g. see
(\textsf{\textbf{V}}) to (\textsf{\textbf{VII}}) for $n=2$). The limitation for
high $n$ is the limited number of auxiliary states available in $^{25}$Mg$^+$.
However, other techniques for phonon counting up to $n=10$ by exploiting trap
induced Kerr-nonlinearities have been demonstrated~\cite{ding_cross-kerr_2017}
and modifications using laser-induced Kerr nonlinearities~\cite{Stobinska_generation_2011} combined with continuous dynamic decoupling
techniques~\cite{bermudez_robust_2012} might be an option for future
implementations.

\section{Quantum metrology}
The precision of an estimation is bounded by means of the Cram\'er-Rao bound as
\begin{align}
  \Delta\theta_{\mathrm{est}}\geq \Delta\theta_{\mathrm{CR}}=\frac{1}{\sqrt{N\mathcal{F}(\theta)}},
\end{align}
where $\theta_{\mathrm{est}}$ is an arbitrary estimator for $\theta$, $N$ is the number of repeated measurements, and 
\begin{align}
  \mathcal{F}(\theta)=\sum_{\mu}\frac{1}{P(\mu|\theta)}\left(\frac{\partial P(\mu|\theta)}{\partial \theta}\right)^2
\end{align}
is the (classical) Fisher information. The probability distribution $P(\mu|\theta)=\mathrm{Tr}\{\hat{\Pi}_{\mu}\hat{\rho}(\theta)\}$ is determined by the quantum state $\hat{\rho}(\theta)$ and the choice of measurement, described by the projectors $\{\hat{\Pi}_{\mu}\}_{\mu}$. We consider scenarios in which the unknown phase $\theta$ is imprinted by a unitary process, i.e. $\hat{\rho}(\theta)=\hat{U}(\theta)\hat{\rho} \hat{U}(\theta)^{\dagger}$ with $\hat{U}(\theta)=e^{-i\hat{H}\theta}$.

The mean value $\langle \hat{M}\rangle_{\hat{\rho}(\theta)}=\mathrm{Tr}\{\hat{M}\hat{\rho}(\theta)\}$ and variance $(\Delta\hat{M})^2_{\hat{\rho}(\theta)}=\langle \hat{M}^2\rangle_{\hat{\rho}(\theta)}-\langle \hat{M}\rangle_{\hat{\rho}(\theta)}^2$ of the measured observable $\hat{M}=\sum_{\mu}\mu\hat{\Pi}_{\mu}$ can be used to derive a lower bound for the Fisher information ~\cite{pezze_quantum_2014}
\begin{align}
  \mathcal{F}(\theta)\geq \frac{1}{(\Delta\hat{M})^2_{\hat{\rho}(\theta)}}\left(\frac{d\langle \hat{M}\rangle_{\hat{\rho}(\theta)}}{d\theta}\right)^2.
  \label{eq:FishInf}
\end{align}
This bound is tight if there are only the two measurement outcomes $\mu=1,0$ with $P(1|\theta)=1-P(0|\theta)$ and $(\Delta\hat{M})^2_{\hat{\rho}(\theta)}=P(1|\theta)(1-P(0|\theta))$.

Maximizing the Fisher information over all possible measurements leads to the quantum Fisher information ~\cite{braunstein_statistical_1994}
\begin{align}
  \max_{\{\hat{\Pi}_{\mu}\}}\mathcal{F}(\theta)=\mathcal{F}_Q[\hat{\rho},\hat{H}],
\end{align}
which is a function of the initial state $\hat{\rho}$ and the generator $\hat{H}$ of the unitary evolution. We obtain the quantum Cram\'er-Rao bound as the general precision limit for quantum parameter estimation ~\cite{helstrom_quantum_1976}
\begin{align}\label{eq:QCR}
  \Delta\theta_{\mathrm{est}}\geq\Delta\theta_{\mathrm{CR}}\geq\Delta\theta_{\mathrm{QCR}}=\frac{1}{\sqrt{N\mathcal{F}_Q[\hat{\rho},\hat{H}]}}.
\end{align}

\section{Extracting the Fisher information from experimental data}
We can use the data shown in Fig.~1~\textbf{c} to get a
measured value for the Fisher information of our measurement.  As can be seen from
Eq.~\ref{eq:FishInf}, the Fisher information depends on the slope and the
noise properties of the measurement presented before. The slope $s(\alpha_i)=\frac{d\langle \hat{M}\rangle_{\hat{\rho}(\alpha)}}{d\alpha}$ is experimentally determined for each displacement amplitude $\alpha_i$ by a symmetric difference quotient 
\begin{eqnarray}
  s(\alpha_i)=\frac{P_{\downk} (\alpha_{i+1})- P_{\downk}(\alpha_{i-1})}{\alpha_{i+1}-\alpha_{i-1}}.
\end{eqnarray} 
For the first and last measurement point is is determined by an asymmetric difference quotient
\begin{eqnarray}
  s(\alpha_i)= \frac{P_{\downk} (\alpha_{i+1})- P_{\downk}(\alpha_{i})}{\alpha_{i+1}-\alpha_{i}}
\end{eqnarray}
As discussed before, the noise is dominated by quantum projection noise.

\section{Estimation of the achievable force sensitivity} The Amplitude $F$ of a resonant oscillating force ,
required to get a displacement of $\alpha$ after time $t_{\text{F}}$ is given by
\begin{eqnarray}
  F=\frac{2\hbar}{z_0 t_{\text{F}}}\times \alpha,
\end{eqnarray}
where $z_0=\sqrt{\hbar/2m\omega_z}$ is the ground state wave packet extent for
an atom with mass $m$, trapped in a harmonic potential with trap frequency
$\omega_z$.  Therefore, the statistical uncertainty for a force estimation can
be written as 
\begin{eqnarray} 
 \Delta F& =& \frac{2\hbar}{z_0 t_{\text{F}}}\times\Delta\alpha\\ 
 &\geq&\frac{2\hbar}{z_0 t_{\text{F}}}\frac{1}{\sqrt{\mathcal{F}N}},
\end{eqnarray}
with  $\mathcal{F}$, the Fisher information for the $\alpha$-estimation (measured result shown in
Fig.~2 \textsf{a}) and $N$ the number of experiments.  Rewriting this expression with
$N=\tau/t_{\text{cycle}}$, and introducing $R_F=t_{\text{cycle}}/t_{\text{F}}$ as the
ratio of cycle time and  $t_{\text{F}}$ we obtain
\begin{eqnarray} 
  \frac{\Delta F}{\sqrt{\Delta}}\geq\frac{2\hbar R_F}{z_0}\sqrt{\frac{1}{\mathcal{F}\tau}},
\end{eqnarray} 
where $\Delta=1/t_{\text{cycle}}$ is the measurement bandwidth and $\tau$ the total measurement time.
For $^{25}$Mg$^+$ with $\omega_z=2\pi\times\unit[1.89]{MHz}$, $t_F=\unit[10]{\upmu s}$,
$t_\text{cycle}=\unit[15]{ms}$, $\mathcal{F}=5$ and $\tau=t_\text{cycle}$ the
resulting force sensitivity is $\unit[112]{yN/\sqrt{Hz}}$.

\section{Oscillation amplitude}
For a harmonic oscillator, the position observable $\hat{x}$ is related to the
quadrature component $\hat{X}=\frac{1}{\sqrt{2}}\left(\hat{a}^{\dagger}+\hat{a}\right)$ by
  \begin{equation}
    \hat{z}=\sqrt{\frac{\hbar}{m\omega_z}}\hat{X}
    \label{Eq:PosQuad}
  \end{equation}
From this relation the expectation value of the position operator for a
coherent state $\alpha$ can be evaluated to be
  \begin{equation}
    \langle \hat{z}\rangle_\alpha = \sqrt{\frac{\hbar}{2m\omega_z}}2\alpha\cos{\omega_z
    t}=2z_0\alpha \cos{\omega_z t}\,.
    \label{eq:ExpPos}
  \end{equation}
Therefore the oscillation amplitude for a given displacement is $A=2z_0\alpha$.
Accordingly, the y-axis in figure \ref{Fig:FisherInfo} \textbf{\textsf{b}} was
scaled by $\Delta A=2z_0\Delta\alpha$.

\begin{figure} \centering
  \includegraphics[width=183mm]{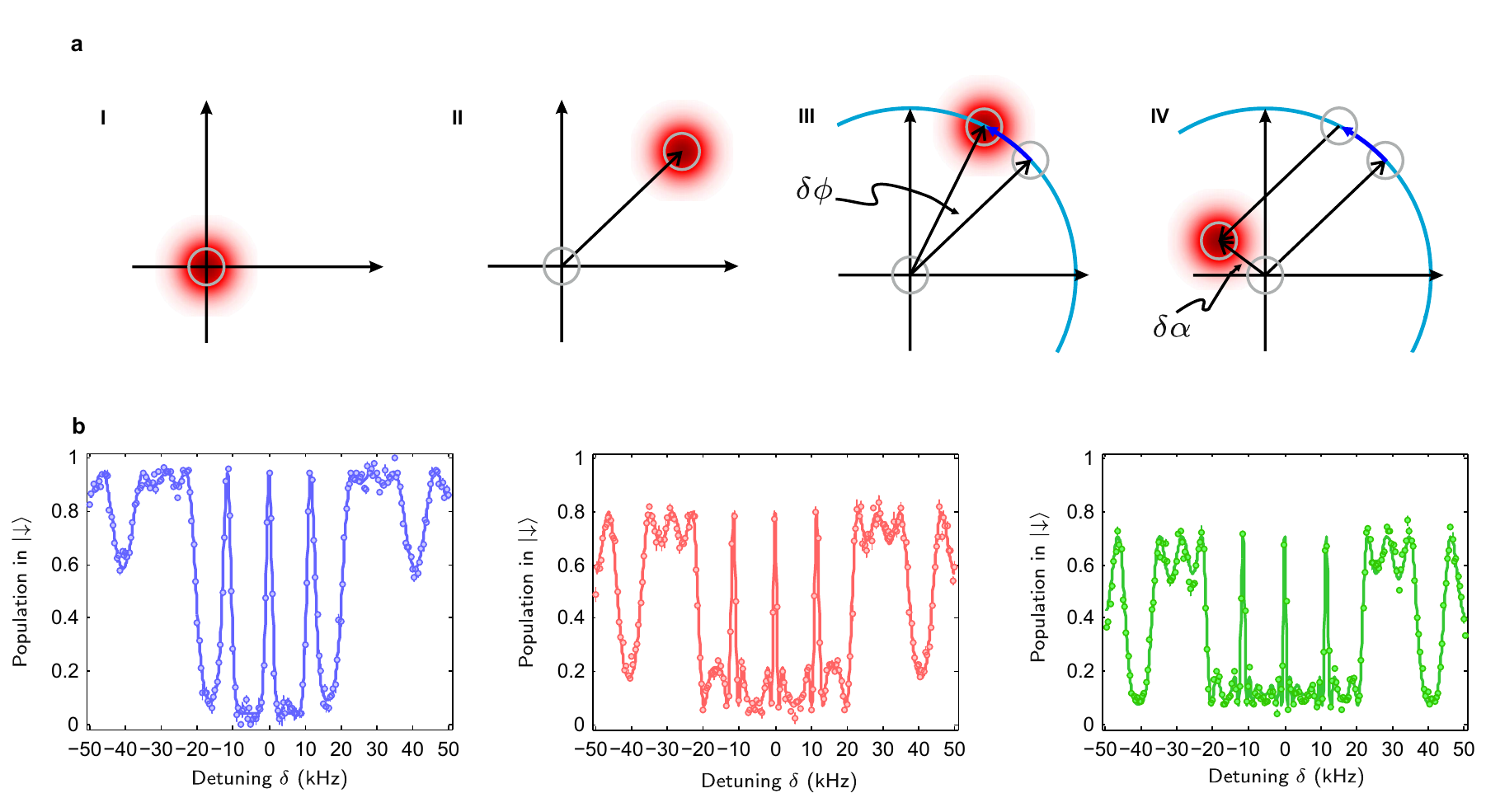}
  \caption{\textbf{Ramsey scheme for trapping frequency measurement }\textbf{\textsf{A}} Ramsey scheme to measure the trap frequency.
  (\textbf{\textsf{I}}) The ion is prepared in the motional ground state (or
  higher Fock state, not shown in Figure). After a displacement implemented with a detuned oscillating force
  (\textbf{\textsf{II}}) the ion's motion accumulates a phase $\phi=\delta\times T$ during a waiting time
  $T$ (\textbf{\textsf{III}}). When the displacement is undone (\textbf{\textsf{IV}}), a
  residual displacement remains, which depends on the accumulated phase.
  Measuring the residual displacement with the overlap measurement technique
  described above for different detuning gives the Ramsey fringes shown in
  subfigure \textsf{\textbf{B}}. The experimental results are shown for three
different Fock states (blue: $n=0$, red: $n=1$, green: $n=2$).  It can be seen,
that with increasing Fock state order, the width of the resonance lines
decreases. Each point is an average of 1250 experiments. The solid line is a fit to the theoretical lineshape (Supplementary Information Eq.~(17)) considering the reduced contrast.}
\label{Fig:RamseyFull} \end{figure} 
\cleardoublepage
\section*{Supplementary information}
\section{Error analysis for Fisher information measurement} 
\begin{table}[b]
  \begin{tabular}{@{}lr@{}} 
 \toprule[1pt]\addlinespace[2pt]
    $\left(\Delta \mathcal{F}_5\right)_{s_\text{QPN}}$&$0.427$\\
    $\left(\Delta \mathcal{F}_5\right)_\text{fit}$&$0.006$\\ 
    $\left(\Delta \mathcal{F}_5\right)_\text{fin}$&$0.464$\\ 
    $\left(\Delta \mathcal{F}_5\right)_\text{QPN}$&$0.068$\\\addlinespace[2pt]\bottomrule[1pt]
  \end{tabular} 
  \caption{error budget for the Fisher information estimation of
  measurement point $i=5$ with $\alpha=0.59$.} \label{tab:errorbudget}
\end{table} Below we will use the shortened notation $ P_{\downk}
(\alpha_{i})\equiv P_i $ and $s(\alpha_i)\equiv s_i$.  
\subsection{Errors on slope determination}
 \paragraph{Quantum projection noise}
The slope is determined by measuring the population at two neighboring
points. The error associated with this measurement is $\Delta
P_{i-1}=\sqrt{\frac{P_{i-1}(1-P_{i-1})}{N}}$ and $\Delta
P_{i+1}=\sqrt{\frac{P_{i-1}(1-P_{i-1})}{N}}$, where N is the number of
independent measurements and is propagated to the slope error due to quantum
projection noise via \begin{eqnarray} \left(\Delta
  s_i\right)_{QPN}&=&\sqrt{\left(\frac{\partial s_i}{\partial P_{i-1}}
  \Delta P_{i-1}\right)^2+\left(\frac{\partial s_i}{\partial P_{i+1}}
  \Delta P_{i+1}\right)^2}\\
  &=&\sqrt{\frac{1}{N}\frac{P_{i-1}(1-P_{i-1})+P_{i+1}(1-P_{i+1})}{\left(\alpha_{i+1}-\alpha_{i-1}\right)^2}}
\end{eqnarray} 
\paragraph{Fit error for displacement determination }
To determine the implemented displacement, we fitted the expected theoretical curve
to the measured data for the motional ground state. The fit error $\Delta
\alpha$ also gives an error on the denominator in the difference quotient. The
resulting slope error reads
\begin{eqnarray} \left(\Delta
  s_i\right)_{fit}=\frac{s_i}{\alpha_{i+1}-\alpha_{i-1}}\Delta{\alpha}
\end{eqnarray}

\paragraph{Error due to finite step size}
The difference quotient only gives an
approximation of the true slope of the signal. For a symmetric derivative the first order
error to this approximation is given by 
\begin{eqnarray} 
  \left(\Delta s_i\right)_{fin}=\left(\frac{\text{d}^3 P}{\text{d}\alpha^3}\right)_i\frac{(\alpha_{i+1}-\alpha_{i-1})^2}{6},
\end{eqnarray} 
and in the asymmetric case by 
\begin{eqnarray} 
  \left(\Delta s_i\right)_{fin}=\left(\frac{\text{d}^2 P}{\text{d}\alpha^2}\right)_i\frac{(\alpha_{i+1}-\alpha_{i-1})}{2},
\end{eqnarray}

For the error estimation we used the theoretically calculated derivative of 
$P=\exp({-|\alpha|^2})\mathcal{L}_n(|\alpha|^2)^2$, where $n$ denotes the number
of excitations in the Fock state, without additional parameters accounting for the reduced contrast observed in the experiment.

\paragraph{Total slope error}
The total error on the estimation of the signal slope $s$ then reads 
\begin{eqnarray} 
  \left(\Delta s_i\right)=\sqrt{\left(\Delta s_i\right)_{QPN}^2+\left(\Delta s_i\right)_{fit}^2+\left(\Delta s_i\right)_{fin}^2} \end{eqnarray}

\subsection{Errors on QPN determination}
Since we use a binary data set to determine the expectation value for the spin measurement, the value for the
variance is exact. However, for finite $N$  statistical fluctuation will give
rise to an uncertainty given by $\Delta QPN_i=P_i(1-P_i)\sqrt{\frac{2}{N-1}}$~~\cite{casella_methods_2001}.

 \subsection{Total error } The total
estimated error on the measurement of the Fisher information is given by
\begin{eqnarray} 
  \Delta \mathcal{F} =\sqrt{\left(\frac{2s}{P_i(1-P_i)}\Delta s_i\right)^2 +\left(\frac{s^2}{(P_i(1-P_i))^2}\Delta QPN_i\right)^2}.
\end{eqnarray} 
The most significant violation of the standard quantum limit has
been observed for point $i=5$ in the Fock state $n=1$ data. The Fisher
Information for this measurement was $\mathcal{F}_5=5.37(63)$ and the different
uncertainties are summarized in the error budget in table \ref{tab:errorbudget}

\section{Ramsey pattern line shape}
The individual displacement pulses start at $t_0$ and are applied for a duration $t_F$. The oscillating force is detuned by $\delta$ from the axial trap frequency $\omega_z$, resulting in the time dependent interaction Hamiltonian
\begin{equation}
    \hat{H} = i \hbar \left( \gamma(t) \hat{a}^{\dagger} - \gamma^{*}(t)\hat{a} \right)
\end{equation}
where $\gamma(t) = i \Omega e^{i (\delta \cdot t + \phi_{LO})} $ and $\Omega = -q E_0 z_0/(2\hbar)$ (see also section \textit{Trap modulation to implement displacement operator}).
The dynamics is given by the unitary evolution ~\cite{roos_ion_2008}
\begin{equation}
    \hat{U}(t_0, t) = \hat{D}\big(\alpha(t_0, t)\big) e^{i \Phi(t_0, t)}
\end{equation}
where the interaction starts at $t_0$ and has duration $t$.
The displacement and phase are:
\begin{align}
    \alpha(t_0, t) &= \int_{t_0}^{t_0+t}\mathrm{d}\tau\, \gamma(\tau)   \\
    \Phi(t_0, t) &= \mathrm{Im} \left[\int_{t_0}^{t_0+t}\mathrm{d}\tau\, \gamma(\tau) \int_{t_0}^{\tau}\mathrm{d}\tau'\, \gamma^{*}(\tau') \right] \,\,.
\end{align}
For the total sequence the evolution is then
\begin{equation}
    \hat{U}_{tot} =\hat{ U}_{\phi_{LO}+\pi}\big(t_0+t_F+T, t_F\big)\, \hat{U}_{\phi_{LO}}\big(t_0, t_F\big)
\end{equation}
where subscripts represent a phase change of the local oscillator to undo the initial displacement.
Up to a global phase this results in a displacement
\begin{equation}
    \hat{U}_{tot} = \hat{D}\left(\delta\alpha\right)
\end{equation}
where  $\delta \alpha=\alpha(t_0, t_F) - \alpha(t_0+t_F+T, t_F) $ with the two contributions
\begin{align}
    \alpha(t_0, t_F) &= \dfrac{\Omega e^{i\phi_{LO}}}{\delta} \left(e^{i\delta(t_0+t_F)} - e^{i\delta t_0} \right) = \dfrac{i\Omega t_F e^{i (\phi_{LO}+\delta t_0)}}{(\delta t_F/2)} \sin \left( \dfrac{\delta t_F}{2} \right) e^{i \delta t_F/2}  \\
    \alpha(t_0+t_F+T, t_F) &= \dfrac{\Omega e^{i\phi_{LO}}}{\delta} \left(e^{i\delta(t_0+ 2 t_F + T)} - e^{i\delta (t_0+t_F+T)} \right) = \dfrac{i\Omega t_F e^{i (\phi_{LO}+\delta t_0)}}{(\delta t_F/2)} \sin \left( \dfrac{\delta t_F}{2} \right) e^{i \delta (3t_F+2T)/2} \, .
\end{align}
So the residual displacement at the end of the sequence is
\begin{equation}
    \delta \alpha = 2 \Omega t_F\, \mathrm{sinc} \left(\dfrac{\delta \, t_F}{2}\right) \sin \left(\dfrac{\delta (T+t_F)}{2}\right) \cdot e^{i \phi_{LO}} e^{i\delta (t_0 +t_F +T/2)}
\end{equation}
This residual displacement can be detected by the overlap measurement described in the main text and gives the final result for the line shape of the Ramsey pattern
\begin{equation}
    \vert \langle n \vert \hat{D}(\delta \alpha) \vert n \rangle \vert^2 = \mathrm{exp}(-\vert \delta \alpha \vert^2) \big(\mathcal{L}_n(\vert \delta \alpha \vert^2)\big)^2\,.
\label{eq:RamseyShape}
\end{equation}

\section{Optimal estimation of a displacement amplitude without phase information}\label{sec:singlemode}
We consider the metrological task of estimating the amplitude of a displacement. The phase of the displacement is unknown at the time of the state preparation. To optimize the sensitivity of the estimation, the `detector' shall be prepared in an optimal quantum state.

The unitary process which generates the phase shift is given by the displacement
\begin{align}\label{eq:disp}
\hat{D}(\alpha)&=\exp\left(\alpha \hat{a}^{\dagger}-\alpha^* \hat{a}\right)\notag\\
&=\exp\left((\hat{a}^{\dagger}e^{-i\phi_\text{LO}}-\hat{a}e^{i\phi_\text{LO}})\frac{\theta}{2}\right)\notag\\
&=\exp\left(-i\hat{R}(\phi_\text{LO})\theta\right),
\end{align}
where we defined real parameters $\theta$ and $\phi_\text{LO}$, such that $\alpha=\theta e^{-i\phi_\text{LO}}/2$ and $\hat{R}(\phi_\text{LO})=(\sin(\phi_\text{LO})\hat{X}+\cos(\phi_\text{LO})\hat{P})/\sqrt{2}$. 
Furthermore, we used $\hat{X}=(\hat{a}+\hat{a}^{\dagger})/\sqrt{2}$ and $\hat{P}=i(\hat{a}^{\dagger}-\hat{a})/\sqrt{2}$. 
Our goal is to estimate the parameter $\theta=2|\alpha|$. 
To distinguish the Fisher Information with respect to $|\alpha|$, $\mathcal{F}$, from the Fisher information with respect to $\theta$, the latter is denoted by a Gothic type $\mathfrak{F}$. The connection between them is
\begin{align}
\mathcal{F}=4\mathfrak{F}\,.
\end{align}
The reason for introducing $\mathfrak{F}$ is to normalize the SQL to one, which is a widely used convention in the literature~~\cite{pezze_non-classical_2016}, whereas in the main manuscript the classical limit is at 4, but the estimated parameter is $|\alpha|$ from the standard definition of the displacment operator.
According to Eq.~(7) of the main text, the sensitivity of the estimation of $\theta$ is bounded by the quantum Fisher information $\mathfrak{F}_Q[\hat{\rho},\hat{R}(\phi_\text{LO})]$, which depends on the phase $\phi_\text{LO}$ via the generator $\hat{R}(\phi_\text{LO})$. In a ``worst-case'' scenario the sensitivity may be reduced to
\begin{align}\label{eq:minF}
  \mathfrak{F}_{\min}[\hat{\rho}]=\min_{\phi_\text{LO}}\mathfrak{F}_Q[\hat{\rho},\hat{R}(\phi_\text{LO})].
\end{align}
In order to prepare the detector such as to render it most sensitive, even in this worst-case scenario, we need to maximize the figure of merit $\mathfrak{F}_{\min}[\hat{\rho}]$. An alternative strategy consists in optimizing the average performance, as quantified by the figure of merit
\begin{align}\label{eq:meanF}
  \mathfrak{F}_{\mathrm{mean}}[\hat{\rho}]=\frac{1}{2\pi}\int_{0}^{2\pi}d\phi \mathfrak{F}_Q[\hat{\rho},\hat{R}(\phi_\text{LO})].
\end{align}
Below, we derive the limits on these two figures of merit as a function of the number of excitations. We will see that a Fock state maximizes the sensitivity in both cases.

\subsection{Optimizing the minimum sensitivity}
Let us first focus on the quantity~(\ref{eq:minF}). Using $ \mathfrak{F}_Q[\hat{\rho},\hat{R}(\phi_\text{LO})]\leq 4(\Delta \hat{R}(\phi_\text{LO}))^2_{\hat{\rho}}$~~\cite{braunstein_statistical_1994}, we find the following upper bound
\begin{align}
  \mathfrak{F}_{\min}[\hat{\rho}]&\leq 4\min_{\phi_\text{LO}} (\Delta \hat{R}(\phi_\text{LO}))^2_{\hat{\rho}}\notag\\
&=2\min_{\mathbf{n}} \mathbf{n}^T\Gamma_{\hat{\rho}}\mathbf{n},\label{eq:mincovmat}
\end{align}
where we introduced the $2\times 2$ covariance matrix
\begin{align}
\Gamma_{\hat{\rho}}=\begin{pmatrix}(\Delta\hat{X})^2_{\hat{\rho}} & \mathrm{Cov}(\hat{X},\hat{P})_{\hat{\rho}}\\
 \mathrm{Cov}(\hat{X},\hat{P})_{\hat{\rho}} & (\Delta\hat{P})^2_{\hat{\rho}}\end{pmatrix},
\end{align}
with
\begin{align}
\mathrm{Cov}(\hat{X},\hat{P})_{\hat{\rho}}&=\frac{1}{2}\mathrm{Tr}\{\hat{\rho}(\hat{X}\hat{P}+\hat{P}\hat{X})\}-\mathrm{Tr}\{\hat{\rho}\hat{X}\}\mathrm{Tr}\{\hat{\rho}\hat{P}\},
\end{align}
and a unit vector $\mathbf{n}=(\sin(\phi_\text{LO}),\cos(\phi_\text{LO}))$. The minimum in Eq.~(\ref{eq:mincovmat}) is given by the smallest eigenvalue $\lambda_{\min}$ of the matrix $\Gamma_{\hat{\rho}}$. This eigenvalue can again be bounded from above:
\begin{align}\label{eq:ineq1}
\lambda_{\min}(\hat{\rho})&=\frac{(\Delta\hat{X})_{\hat{\rho}}^2+(\Delta\hat{P})_{\hat{\rho}}^2}{2}\notag\\
&\quad-\frac{1}{2}\sqrt{((\Delta\hat{X})_{\hat{\rho}}^2-(\Delta\hat{P})_{\hat{\rho}}^2)^2+4\mathrm{Cov}(\hat{X},\hat{P})_{\hat{\rho}}^2}\notag\\
&\leq \frac{1}{2}\mathrm{Tr}\{\hat{\rho}(\hat{X}^2+\hat{P}^2)\}\notag\\
&=\mathrm{Tr}\{\hat{\rho}(\hat{a}^{\dagger}\hat{a}+1)\}\notag\\
&=n+\frac{1}{2},
\end{align}
where $n=\mathrm{Tr}\{\hat{\rho}\hat{a}^{\dagger}\hat{a}\}$ determines the number of excitations. Hence, the minimal sensitivity is generally bounded by
\begin{align}\label{eq:minbound}
  \mathfrak{F}_{\min}[\hat{\rho}]&\leq 2n+1.
\end{align}

\subsection{Optimal states must be quantum non-Gaussian}
States which reach the upper bound~(\ref{eq:minbound}) for $n>0$ must necessarily be quantum non-Gaussian, i.e., they cannot be written as a mixture of Gaussian states (see e.g. Refs.~~\cite{filip_detecting_2011,genoni_detecting_2013}). 

To see this, notice that in~(\ref{eq:ineq1}), equality is reached if and only if the conditions $\langle \hat{X}\rangle_{\hat{\rho}}=\langle\hat{P}\rangle_{\hat{\rho}}=\mathrm{Cov}(\hat{X},\hat{P})_{\hat{\rho}}=0$ and $\Delta\hat{X}_{\hat{\rho}}=\Delta\hat{P}_{\hat{\rho}}$ are satisfied, or equivalently, $\langle \hat{a}\rangle_{\hat{\rho}}=\langle \hat{a}^{\dagger}\rangle_{\hat{\rho}}=\langle \hat{a}\hat{a}\rangle_{\hat{\rho}}=\langle \hat{a}^{\dagger}\hat{a}^{\dagger}\rangle_{\hat{\rho}}=0$. These conditions can only be satisfied by a Gaussian state when $n=0$, i.e., the vacuum state $|0\rangle$. The above statement then follows together with the convexity of the quantum Fisher information.

All pure states lead to equality in~(\ref{eq:mincovmat}). In the case of a pure state, non-Gaussianity is equivalent to a negative Wigner function ~\cite{hudson_when_1974}.

Hence, mixtures of Gaussian states will always perform sub-optimally for the estimation of a displacement amplitude in a ``worst-case'' scenario.

\subsection{Average sensitivity bound}
The same bound also holds for the average~(\ref{eq:meanF}). We obtain
\begin{align}\label{eq:meanbound}
  \mathfrak{F}_{\mathrm{mean}}[\hat{\rho}]\leq (\Delta\hat{X})_{\hat{\rho}}^2+(\Delta\hat{P})_{\hat{\rho}}^2\leq 2n+1,
\end{align}
and equality is reached only by states that are not displaced: $\langle \hat{X}\rangle_{\hat{\rho}}=\langle\hat{P}\rangle_{\hat{\rho}}=0$.

\subsection{Optimality of Fock states}\label{sec:optfockstates}
The bounds~(\ref{eq:minbound}) and~(\ref{eq:meanbound}) are saturated by Fock states $|n\rangle=(\hat{a}^{\dagger})^n/\sqrt{n!}|0\rangle$.  Indeed, one easily confirms that Fock states satisfy all of the optimality conditions for the two bounds~(\ref{eq:minbound}) and~(\ref{eq:meanbound}), as stated above. Specifically, one obtains $(\Delta\hat{X})_{|n\rangle}^2=\langle n|\hat{X}^2|n\rangle=\frac{1}{2}\langle n|(\hat{a}+\hat{a}^{\dagger})^2|n\rangle=n+\frac{1}{2}$, and similarly for $(\Delta\hat{P})_{|n\rangle}^2$, as well as $\mathrm{Cov}(\hat{X},\hat{P})_{|n\rangle}=0$. This yields the covariance matrix
\begin{align}\label{eq:covfock}
\Gamma_{|n\rangle}=\begin{pmatrix} n+\frac{1}{2} & 0\\ 0 & n+\frac{1}{2}\end{pmatrix},
\end{align}
with the two-fold degenerate eigenvalue $n+1/2$ (the degeneracy expresses the fact that the Fock state is equally sensitive in all directions). Since $\mathfrak{F}_Q[|\Psi\rangle,\hat{R}(\phi_\text{LO})]=4 (\Delta\hat{R}(\phi_\text{LO}))^2_{|\Psi\rangle}$ for all pure states, we find the exact equality
\begin{align}
  \mathfrak{F}_{\min}[|n\rangle]=2n+1.
\end{align}

We obtain the same result for the mean value~(\ref{eq:meanF}), since for pure states, the equality
\begin{align}
  \mathfrak{F}_{\mathrm{mean}}[|\Psi\rangle]&=\mathrm{Tr}\Gamma_{|\Psi\rangle},
\end{align}
holds. For the Fock state $|n\rangle$ this leads to $\mathfrak{F}_{\mathrm{mean}}[|n\rangle]=2n+1$.

In summary, for a fixed energy (given by $n$), the Fock state provides the optimal precision for displacement detections with unknown phase. This is true for both strategies, i.e., preparing for the worst-case scenario~(\ref{eq:minF}) or optimizing the mean performance~(\ref{eq:meanF}).
\section{Classical limit}\label{sec:P}
Based on the Glauber-Sudarshan P-representation,
\begin{align}\label{eq:cl}
\hat{\rho}=\int d\alpha P(\alpha)|\alpha\rangle\langle\alpha|,
\end{align}
we define classical states $\hat{\rho}_{\mathrm{cl}}$ as those for which $P(\alpha)$ describes a probability distribution. Here $|\alpha\rangle=\hat{D}(\alpha)|0\rangle$ is a coherent state. The classical limit is then given as the maximum quantum Fisher information, taken over all classical states $\hat{\rho}_c$. Since the Fisher information is convex, the maximum is attained by a pure coherent state $|\alpha\rangle$. Using $\mathfrak{F}_Q[|\alpha\rangle,\hat{R}(\phi_\text{LO})]=4 (\Delta\hat{R}(\phi_\text{LO}))^2_{|\alpha\rangle}$, we obtain
\begin{align}\label{eq:cllimit}
  \mathfrak{F}_Q^{(\mathrm{cl})}[\hat{R}(\phi_\text{LO})]:=\max_{\hat{\rho}_{\mathrm{cl}}}\mathfrak{F}_Q[\hat{\rho}_{\mathrm{cl}},\hat{R}(\phi_\text{LO})]=1,
\end{align}
which is independent of $\phi$ and $\alpha$ and corresponds to the sensitivity of the vacuum state. Thus, any observation of
\begin{align}
  \mathfrak{F}_Q[\hat{\rho},\hat{R}(\phi_\text{LO})]>1,
\end{align}
reveals that the state $\hat{\rho}$ is non-classical according to the above definition ~\cite{rivas_precision_2010}. This form of non-classicality is a necessary resource to overcome the classical limit eq.~(\ref{eq:cllimit}).
The resulting upper bound on the Fisher information for amplitude measurements using classical states is given by 
\begin{align}
  \mathcal{F}_Q=4.
\end{align}
The corresponding bound for the phase measurement can be infered from
\begin{align}
\mathcal{F}_Q(\delta)=\mathcal{F}_Q(\alpha)\left(\frac{\mathrm{d(\delta\alpha)}}{\mathrm{d\alpha}}\right)^2\,.
\end{align}
Assuming that the detuning $\delta$ is small compared to the displacement $\Omega$ rate, we get
\begin{align}
\delta \alpha = \Omega t_F \left(T+t_F\right)\delta+\mathcal{O}(\delta^3)
\end{align}
and for the Quantum Fisher information
\begin{align}
\mathcal{F}_Q(\delta)=\left(2|\alpha|\left(T+t_f\right) \right)^2\,,
\end{align}
which results in the SQL limit given in eq.~(3) in the main text.
\section{Analogy with a two-mode interferometer}
\begin{figure}
\centering
\includegraphics[width=89mm]{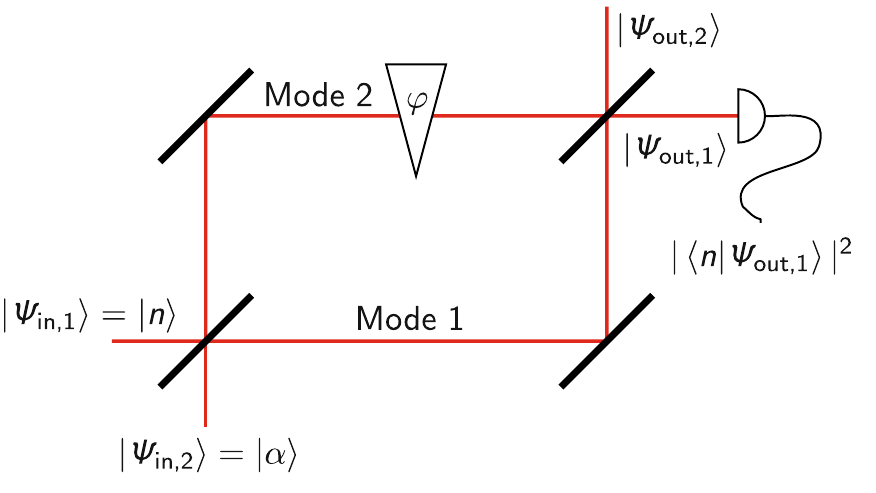}
\caption{\textbf{Analogy with a two-mode interferometer.} The presented measurements can be understood in terms of a two-mode interferometer, with a coherent state and a Fock state at the input ports $\ket{\psi_\text{in}}=\ket{n}\ket{\alpha}$. The output state is $\ket{\Psi_\text{out}}=\hat{U}(\varphi)\ket{\Psi_\text{in}}$, with $\hat{U}$ given in Eq.~\ref{eq:MZ}}
\end{figure}
Fock-state metrology can be understood in the wider context of a two-mode Mach-Zehnder interferometer ~\cite{pezze_ultrasensitive_2013}. The evolution in a two-mode interferometer (with modes $\hat{a}$ and $\hat{b}$) is described by,
\begin{align}\label{eq:MZ}
\hat{U}(\varphi)=\exp\left(-i\hat{J}_y\varphi\right),
\end{align}
where $\hat{J}_y=(\hat{a}^{\dagger}\hat{b}-\hat{b}^{\dagger}\hat{a})/2i$ and $\varphi$ is the phase shift. 

For the two-mode transformation~(\ref{eq:MZ}) one obtains the classical limit, i.e., the maximum sensitivity for two-mode classical states, 
\begin{align}\label{eq:shotnoise}
  \mathfrak{F}_Q^{(\mathrm{cl})}[\hat{J}_y]=n_a+n_b,
\end{align}
where $n_a+n_b$ is the total number of excitations ($n_a=\mathrm{Tr}\{\hat{\rho} \hat{a}^{\dagger}\hat{a}\}$ and $n_b=\mathrm{Tr}\{\hat{\rho} \hat{b}^{\dagger}\hat{b}\}$). This bound is known as the shot-noise limit and coincides with the sensitivity bound for states that are separable among particles ~\cite{pezze_entanglement_2009}. 
For a Fock state $|n\rangle$ in input mode $\hat{a}$, the quantum Fisher information reads 
\begin{align}\label{eq:QFIFOCK}
  \mathfrak{F}[|n\rangle\langle n|\otimes \hat{\rho}_b,\hat{J}_y]=2n_bn+n_b+n,
\end{align}
which yields a quantum-enhanced sensitivity for any $n>0$ ~\cite{pezze_ultrasensitive_2013}, in agreement with the results from the previous section.

The specific case of the displacement discussed in Sec.~\textbf{OPTIMAL ESTIMATION OF A DISPLACEMENT AMPLITUDE WITHOUT PHASE INFORMATION} is recovered in the homodyne limit, in which a classical, highly populated coherent state is inserted in one of the input ports. We thus assume that input mode $\hat{b}$ is prepared in a coherent state $\hat{\rho}_b=|\alpha_0\rangle\langle\alpha_0|$ with $|\alpha_0|^2=n_0\gg 1$ particles, while the quantum state $\hat{\rho}_a$ of mode $\hat{a}$ input is arbitrary. Neglecting quantum fluctuations, by making the replacement $\hat{b}\rightarrow\alpha$ and $\hat{b}^{\dagger}\rightarrow \alpha^*$ in Eq.~(\ref{eq:MZ}), we obtain an effective transformation of mode $\hat{a}$, described by
\begin{align}
\hat{U}(\theta)=\exp\left(-(\alpha_0\hat{a}^{\dagger}-\alpha_0^{*}\hat{a})\varphi/2\right).
\end{align}
The evolution of mode $\hat{a}$ is effectively given by the displacement~(\ref{eq:disp}) if we rescale the parameter by $\alpha=-\alpha_0\varphi/2$. The precision of an estimation of the phase shift $\phi$ in the two-mode Mach-Zehnder interferometer is bounded by $\mathfrak{F}_Q[\hat{\rho},\hat{H}(\phi_\text{LO})]$, where $\hat{H}(\phi_\text{LO})=-\sqrt{n_0}\hat{R}(\phi_\text{LO})$. The precision limits can be linked to the single-mode results from Sections~\textbf{QUANTUM METROLOGY} and~\textbf{CLASSICAL LIMIT} by a simple rescaling transformation: $\mathfrak{F}_Q[\hat{\rho},\hat{H}(\phi_\text{LO})]=n_0\mathfrak{F}_Q[\hat{\rho},\hat{R}(\phi_\text{LO})]$. Hence, the non-classicality bound for the Mach-Zehnder interferometer in the homodyne limit can be obtained from Eq.~(\ref{eq:cllimit}) as $\mathfrak{F}^{(\mathrm{cl})}_Q[\hat{H}(\phi_\text{LO})]=n_0$. For a Fock state in the second input port, we obtain a precision of $\mathfrak{F}_Q[||n\rangle,\hat{H}(\phi_\text{LO})]=2n_0n+n_0$, independently of the phase of $\alpha_0$, which allows for sub-shot-noise sensitivity for $n>0$, in agreement with our previous considerations (see Sec.~\textbf{OPTIMAL ESTIMATION OF A DISPLACEMENT AMPLITUDE WITHOUT PHASE INFORMATION}).

As discussed above, the classical bound and the Fock-state sensitivity can be equivalently obtained from the two-mode results~(\ref{eq:shotnoise}) and~(\ref{eq:QFIFOCK}), remembering that in the homodyne limit considered here, the contribution of $n_a$ or $n$ to the total number of particles is negligible due to $n_b=n_0\gg n_a,n$.
\vspace{1cm}

\end{document}